\documentclass[aps,prd,twocolumn,groupedaddress]{revtex4-1}
\usepackage{graphicx}
\include{amssym}

\begin{document}
\title{Low-energy theorem for $\gamma\to 3\pi$: surface terms against $\pi a_1$-mixing}

\author{A. A. Osipov$^{1}$}
\email[]{aaosipov@jinr.ru}

\author{M. M. Khalifa$^{2,3}$}
\email[]{mkhalifa@phystech.edu}

\author{B. Hiller$^{4}$}
\email[]{brigitte@fis.uc.pt}

\affiliation{$^1$Bogoliubov Laboratory of Theoretical Physics, Joint Institute for Nuclear Research, Dubna, 141980, Russia}
\affiliation{$^2$Moscow Institute of Physics and Technology, Dolgoprudny, Moscow Region 141701, Russia}
\affiliation{$^3$Department of Physics, Al-Azhar University, Cairo 11751, Egypt}
\affiliation{$^4$CFisUC, Department of Physics, University of Coimbra, P-3004-516 Coimbra, Portugal}

\begin{abstract} 
We reconsider the contribution due to $\pi a_1$-mixing to the anomalous $\gamma\to\pi^+\pi^0\pi^-$ amplitude from the standpoint of the low-energy theorem $F^{\pi}=e f_\pi^2 F^{3\pi}$, which relates the electromagnetic form factor $F_{\pi^0\to\gamma\gamma}=F^\pi$ with the form factor $F_{\gamma\to\pi^+\pi^0\pi^-}=F^{3\pi}$ both taken at vanishing momenta of mesons. Our approach is based on a recently proposed covariant diagonalization of $\pi a_1$-mixing within a standard effective QCD-inspired meson Lagrangian obtained in the framework of the Nambu-Jona-Lasinio model. We show that the two surface terms appearing in the calculation of the anomalous triangle quark diagrams or AVV- and AAA-type amplitudes are uniquely fixed by this theorem. As a result, both form factors $F^\pi$ and $F^{3\pi}$ are not affected by the $\pi a_1$-mixing, but the concept of vector meson dominance (VMD) fails for $\gamma\to\pi^+\pi^0\pi^-$. 
\end{abstract}

\maketitle

\section{Introduction}
The Wess-Zumino \cite{Wess71} effective action precisely describes all effects of QCD anomalies in low-energy processes with photons and Goldstone bosons. The topological content of this action was clarified by Witten \cite{Witten83}. The extension to the case with spin-1 mesons is not unique, and has been discussed in different frameworks. In the massive Yang-Mills approach this has been done in \cite{Schechter84} by gauging the chiral $U(3)\times U(3)$ group. This forces one to choose Bardeen's form of the anomaly \cite{Bardeen69}, that explicitly breaks the global chiral $U(3)\times U(3)$ symmetry. The breaking survives even if the external gauge fields are absent. To get around this difficulty Fujiwara et al. \cite{Fujiwara85} used a framework where vector mesons were identified with dynamical gauge bosons of the hidden local $U(3)_V$ symmetry. This approach does not change the Wess-Zumino action, which now gets an additional anomaly-free term with vector mesons. It represents a homogeneous solution of the inhomogeneous linear differential equation known as the Wess-Zumino consistency condition \cite{Wess71}. As the Wess-Zumino action, this term does not preserve the intrinsic parity [The intrinsic parity of a particle is $+1$ if it transforms as a true Lorentz tensor and is $-1$ for a pseudotensor]. This approach has been generalized to include the axial vector mesons in  \cite{Kaiser90}, where the fourteen independent terms with a priori unknown (real) coefficients totally parameterized the structure of the homogeneous solution. 

There are troublesome questions which arise as soon as one includes the spin-1 states to the effective action. One of them is related to vector meson dominance. The Wess-Zumino action gives correct predictions for a set of low-energy processes, e.g., $\pi^0\to \gamma\gamma$, $\gamma\to 3\pi$ without any reference to the massive vector mesons. If one includes these states one should demonstrate how VMD is possible. In particular, it has been shown in \cite{Fujiwara85} that the complete VMD is not valid in either $\pi^0\to \gamma\gamma$ or $\gamma\to 3\pi$ process. 

The other question is about pseudoscalar -- axial-vector mixing ($\pi a_1$-mixing) of meson states. If one includes axial vector mesons this mixing affects the hadronic amplitudes \cite{Gasiorovicz69,Osipov85}. Therefore, one should demonstrate how $\pi a_1$-mixing does not change the predictions of the Wess-Zumino action. This is not a trivial task. In particular, in \cite{Wakamatsu89} it has been reported that in  a number of well-known models \cite{Schwinger67,Wess67,Sakurai69,Ebert83,r14,Volkov86,Ebert86,Bando85a,Bando85b,Bando85c} the $\pi a_1$-mixing breaks low-energy theorems for some anomalous processes, e.g., $\gamma\to 3\pi$, $K^+K^-\to 3\pi$. The anomalous action derived by Kaiser and Mei\ss ner \cite{Kaiser90} is free from the $\pi a_1$-mixing effects by construction. Nonetheless, it would be instructive to see the mechanism of such  suppression in the pertinent hadron models.

In this paper we address both of the above mentioned issues. For that we derive the low-energy amplitudes $\pi^0\to\gamma\gamma$ and $\gamma\to 3\pi$ in the framework the Nambu-Jona-Lasinio (NJL) model with spin-1 states \cite{Ebert83,r14,Volkov86,Ebert86,Volkov94,Osipov94,Osipov96}. Then we show how the unwanted contributions due to $\pi a_1$-mixing can be suppressed in the $\gamma\to 3\pi$ amplitude. The procedure is based on a careful treatment of the surface terms arising due to the superficial linear divergence of the AVV and AAA triangle graphs (A: axial-vector, V: vector) \cite{Bell69,Jackiw72,Jackiw00}. One should emphasize that the corresponding low-energy theorem of current-algebra \cite{Adler71,Terentiev71,Zee72} 
\begin{equation}
\label{leth}
F^{\pi}=e f_\pi^2 F^{3\pi}
\end{equation} 
can be fulfilled in the NJL model with spin-1 mesons only if there is a deviation from the VMD hypothesis. We come to this conclusion through the gauge covariant treatment of $\pi a_1$-mixing, only recently addressed \cite{Osipov18a,Osipov18b,Osipov18c,Osipov19}. 

\section{The $\pi a_1$-mixing in the $\pi^0\to \gamma\gamma$ decay}
\begin{figure}
\resizebox{0.30\textwidth}{!}{%
  \includegraphics{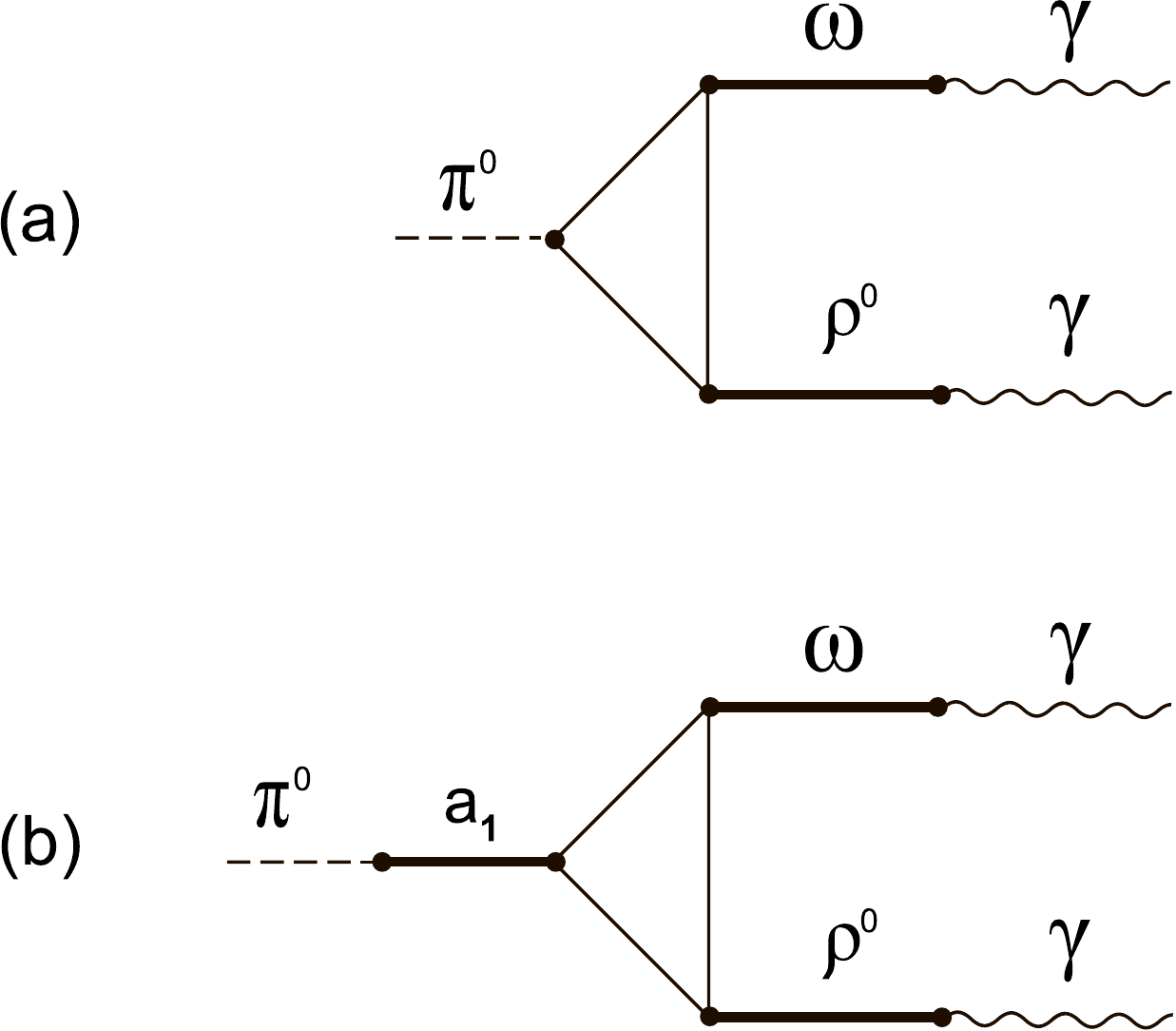}
}
\caption{The two possible graphs for the $\pi^0\to\gamma\gamma$ decay in the NJL model with vector meson dominance.}
\label{fig1}      
\end{figure}

We start with a brief review of the $\pi^0\to\gamma\gamma$ decay in the context of the $\pi a_1$-mixing. This process can be solely described by the VMD-type graph shown in Fig.~\ref{fig1}a. Its contribution is associated with the Lagrangian density \cite{Wess71,Witten83}
\begin{equation}
\label{pigg}
\mathcal{L}_{\pi\gamma\gamma}=-\frac{1}{8}F^\pi\pi^0 e^{\mu\nu\alpha\beta} F_{\mu\nu}F_{\alpha\beta}, \quad F^{\pi}=\frac{N_c e^2}{12\pi^2 f_\pi},
\end{equation}
where $e$ is the electric charge, $F_{\mu\nu}=\partial_\mu A_\nu -\partial_\nu A_\mu$ stands for the strength of the electromagnetic field, $N_c$ is the number of quark colors, and $f_\pi =93\,\mbox{MeV}$ denotes the pion weak decay constant. Let us recall that in the NJL model, one can always switch to spin-1 variables without direct photon-quark coupling, as described in the VMD picture. Then $\mathcal{L}_{\pi\gamma\gamma}$ follows from the direct calculation of the $\pi^0\omega\rho$ quark triangle at leading order of derivative expansion. This yields the current-algebra result $\Gamma (\pi^0\to\gamma\gamma)=7.1 \, \mbox{eV}$ which nicely agrees with the experimental value of $7.9 \, \mbox{eV}$.  

However, in the framework of effective Lagrangians such as  the massive Yang-Mills model \cite{Schechter84}, the hidden symmetry model \cite{Bando85a,Bando85b,Bando85c}, or the NJL model \cite{Ebert83,r14,Volkov86,Ebert86}, there is an additional diagram due to the $\pi a_1$-mixing (see Fig.~\ref{fig1}b). In the NJL model considered here, the latter diagram is an anomalous quark-loop amplitude with axial-vector -- vector -- vector (AVV) vertices. The contribution of this triangle is given by the one-loop integral $\Gamma^{\sigma\mu\nu}(q,p)$, where $q$ and $p$ are the outgoing 4-momenta of $\omega$ and $\rho$ vector mesons, and $\sigma ,\mu ,\nu$ are the Lorentz indices summed with the $a_1$, $\omega$ and $\rho$ polarization vectors, correspondingly.  

As is well-known \cite{Bell69,Jackiw72,Jackiw00}, the evaluation of $\Gamma^{\sigma\mu\nu}(q,p)$ yields a finite answer, although the graph is superficially linearly divergent. Owing to this linear divergence, shifting the integration momentum $k_\alpha\to k_\alpha +a_\alpha$ in the closed quark loop changes the value of the integral, so that there is an essential ambiguity in the linear (in momenta of outgoing particles) part of the loop-function $\Gamma^{\sigma\mu\nu}(q,p)$
\begin{equation}
\label{surf}
\Gamma^{\sigma\mu\nu}(q,p)=-i\frac{N_c g_\rho^3}{16\pi^2}e^{\sigma\mu\nu\alpha} (a+p-q)_{\alpha}+\ldots ,
\end{equation}    
where $g_\rho\simeq \sqrt{12\pi}$ is the coupling of the $\rho\to \pi\pi$ decay. The dots correspond to the contributions of cubic and higher orders in momenta, which are well defined but are not important for our analysis here (the current-algebra theorems are exact to lowest order in momenta). The arbitrary four-momentum $a_\alpha$ can be written, most generally, as a linear combination $a_\alpha =(c_1-1) p_\alpha +(c_2+1) q_\alpha$ with two dimensionless constants $c_1$ and $c_2$, controlling the magnitude of this local part. In the case of the $\pi^0\to\gamma\gamma$ decay, one can fix completely these constants by making use of the vector Ward identities. Indeed, due to the VMD induced transitions $\omega\to\gamma$ and $\rho^0\to\gamma$ the conservation of the electromagnetic current is mandatory in this process. Requiring transversality of $\Gamma^{\sigma\mu\nu}(q,p)$ in each of the two vector indices  
\begin{equation}
\label{trans}
q_\mu\Gamma^{\sigma\mu\nu}(q,p)=0, \quad p_\nu\Gamma^{\sigma\mu\nu}(q,p)=0,
\end{equation} 
one finds $a_\alpha=q_\alpha-p_\alpha$. This means that the AVV triangle of Fig.~\ref{fig1}b does not contribute at leading order of the derivative expansion to the amplitude $\pi^0\to\gamma\gamma$. 

The other aspect of this result is related to the Landau-Yang theorem \cite{Landau48,Yang56} which states that a massive unit spin particle cannot decay into two on shell massless photons. In particular, the theorem forbids the $a_1\to\gamma\gamma$ decay. As a consequence, the axial-vector channel $\pi^0\to a_1\to \gamma\gamma$ induced by the $\pi a_1$-mixing is also forbidden. 

Let us  discuss the issue in terms of the effective Lagrangian describing the hadronic $a_1\omega\rho$ vertex. 
Below we present the result which takes into account the leading and next to the leading orders in the expansion of the AVV quark triangle $\Gamma^{\sigma\mu\nu}(q,p)$ in powers of $q$ and $p$ 
\begin{eqnarray}
\label{aor}
&&\mathcal{L}_{a_1\omega\rho} =\frac{N_c g_\rho^3}{32\pi^2} e^{\sigma\mu\nu\alpha} \left\{a_{1\sigma}^i\left(c_1\omega_\mu \rho^i_{\alpha\nu}+c_2 \rho^i_\nu\omega_{\alpha\mu}\right)  \right. \\
&&\left. -\frac{1}{2m^2}\left[\rho^i_{\alpha\beta} \left(\omega_{\sigma\nu} a^i_{1\beta\mu} + \omega_{\beta\mu}  a^i_{1 \sigma\nu}\right)+2\rho^i_{\sigma\nu}a^i_{1\mu} \partial_\beta\omega_{\beta\alpha} \right]\right\}.\nonumber 
\end{eqnarray} 
Here $b_{\mu\nu} =\partial_\mu b_\nu -\partial_\nu b_\mu$, where $b=\omega, \rho^i, a_1^i$ are spin-1 meson fields, $m$ is the constituent quark mass, and the summation over repeated isospin index $i$ is assumed. The constants $c_1$ and $c_2$ were defined above. Notice, that they represent the freedom related with the surface term (\ref{surf}) appearing in the calculation of the quark AVV triangle. For this reason their values are not intrinsic to the triangle graph, but depend on the context in which they arise \cite{Jackiw00,Baeta01}. For instance, when both vector $\omega$ and $\rho$ mesons couple to photons the gauge symmetry is conserved if and only if $c_1=c_2=0$. On the other hand, in the case of the $a_1\to\gamma\rho$ decay one should preserve transversality of the $\omega\to \gamma$ index and abandon transversality by the Lorentz index related with the $\rho$ field, i.e. the choice is $c_1=0, c_2\neq 0$. A similar argument gives $c_1\neq 0, c_2=0$ for the $a_1\to\gamma\omega$ decay [The three-derivative part of (\ref{aor}) has been used in \cite{Volkov84} to estimate the widths of radiative decays $\Gamma (a_1\to\gamma\rho )=34\,\mbox{keV}$ and $\Gamma (a_1\to\gamma\omega )=300\,\mbox{keV}$. The future phenomenological data should clarify the role of surface terms in these decays]. If one enforces the conservation of the axial-vector current in the AVV-triangle, one finds $c_1=c_2$, as it takes place, for instance, in \cite{Kaiser90}. In the latter case the contribution of the diagram Fig.~\ref{fig1}b vanishes due to the occurrence of an accidental antisymmetry 
under the exchange of fields $\omega_\mu \leftrightarrow \rho^0_\mu$.  

From all the previous considerations we conclude that it is generally most appropriate to use the hadron vertex $a_1\omega\rho$ in the form (\ref{aor}), where the two parameters $c_1$ and $c_2$ should be subsequently specified. The ambiguity contained in the $a_1\omega\rho$ vertex should not scare the reader, because there is no a priori physical process associated with these three particles from which one could extract the values of $c_1$ and $c_2$. Nonetheless, these parameters can be fixed on theoretical or/and phenomenological grounds when the vertex (\ref{aor}) is an element of the Feynman diagram corresponding to a real physical process. In the next section we show how it works for the $\omega\to 3\pi$ amplitude. 

\section{The $\pi a_1$-mixing in the $\omega\to 3\pi$ decay}
Now that we have demonstrated that $\pi a_1$-mixing does not affect the $\pi^0\to\gamma\gamma$ amplitude, and have established the most general low energy structure of the $a_1\omega\rho$ vertex, we can address the main subject of this paper -- the problem of $\pi a_1$-mixing in the $\omega\to 3\pi$ and $\gamma\to 3\pi$ amplitudes. This question has been studied by Wakamatsu \cite{Wakamatsu89} in detail. He has found that the amplitude of the $\omega\to 3\pi$ decay contains uncompensated contributions generated by $\pi a_1$-mixing. This breaks the low energy theorem at the order of $1/a^2$, where 
\begin{equation}
a=\frac{m_\rho^2}{g_\rho^2 f_\pi^2}=1.84
\end{equation}
and $m_\rho =775.26\pm 0.25\, \mbox{MeV}$ is the empirical mass of the $\rho$-meson. Obviously, this conclusion is based on the assumption that VMD is valid. 

\begin{figure}
\resizebox{0.30\textwidth}{!}{%
  \includegraphics{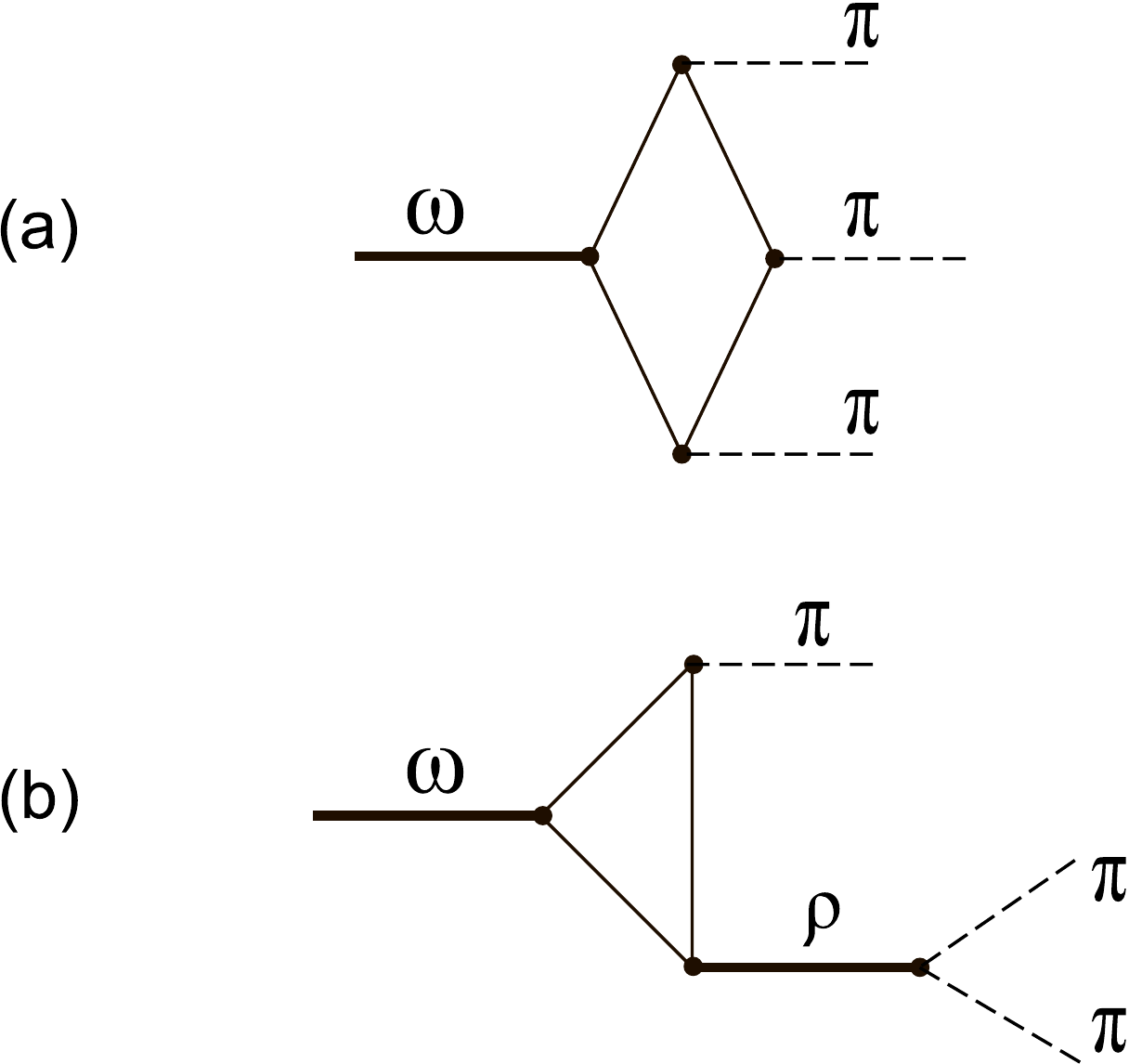}
}
\caption{The quark loop graphs contributing to the $\omega\to 3\pi$ decay in the NJL model. The graph (a) represents a full set of possible diagrams without and with one, two, and three $\pi a_1$-mixing effects on the pion line. The graph (b) represents the diagrams without and with one $\pi a_1$ transitions. The graph with two $\pi a_1$ transitions in the vertex $\rho\to\pi\pi$ is neglected because it contributes to the amplitude only at the next order of derivative expansion.}
\label{fig2}      
\end{figure} 
 
First, let us recall and complement the calculations made in \cite{Wakamatsu89}. The diagrams contributing to the $\omega\to 3\pi$ decay are shown in Fig.~\ref{fig2}, where we have additionally included the box diagram with three $\pi a_1$-transitions and took into account the contribution of the $\omega\rho (a_1\to\pi )$ vertex in the $\rho$-exchange graph neglected in \cite{Wakamatsu89}. The corresponding amplitude is  given by
\begin{equation}   
\label{om3pi} 
A_{\omega\to 3\pi}=-\frac{N_c g_\rho}{4\pi^2 f_\pi^3} e_{\mu\nu\alpha\beta} \epsilon^\mu (q) p_0^\nu p_+^\alpha p_-^\beta 
F_{\omega\to 3\pi},
\end{equation}
where $p_0, p_+, p_-$ denote the momenta of the three pions, $\epsilon^\mu (q)$ is the polarization of the $\omega$-meson with the momentum $q$, and the form factor $F_{\omega\to 3\pi}$ is found to be
\begin{eqnarray}
\label{ff}
F_{\omega\to 3\pi}&=&\left(1-\frac{3}{a}+\frac{3}{2a^2}+\frac{1}{8a^3}\right) \nonumber \\
&+& \left(1-\frac{c}{2a}\right)\sum_{k=0,+,-} \frac{g_\rho^2 f_\pi^2}{m_\rho^2-(q-p_k)^2}.
\end{eqnarray}
Here, in the first parentheses, the contributions of box diagrams without, with one, two, and three $\pi a_1$-transitions are given correspondingly. The last term represents the contribution of two $\rho$-exchange graphs, where $c=c_1-c_2$ controls the magnitude of an arbitrary local part of the AVV-quark-triangle. 

In the low-energy limit, the sum is approximated by the factor $3/a$, arising if one neglects the dependence on momenta in (\ref{ff}). Then one can see that there is a full cancellation among the terms of order $1/a$. This is a well-known result of \cite{Wakamatsu89}. The surface term contributes at order of $1/a^2$. Without this contribution ($c_1=c_2$) we reproduce the $\pi a_1$-mixing effect found in \cite{Wakamatsu89}. 

Heuristically one might have thought that the retention of the surface term in the AVV vertex can be used to cancel the $\pi a_1$-mixing effect. This might be reached by demanding that $c=1+1/(12a)$. However, this 
naive theoretical reasoning is not supported from the phenomenological point of view. For $c=1$, the estimate $\Gamma (\omega\to\pi^+\pi^0\pi^-)=3.2\,\mbox{MeV}$ is too low compared with the well-known experimental value $\Gamma (\omega\to\pi^+\pi^0\pi^-)=7.57\pm 0.13 \,\mbox{MeV}$.

Actually, there is a solid theoretical fact established by Cohen \cite{Cohen89}. He has shown that the chiral Ward identities for the $\gamma\to 3\pi$ process imply that both the chiral triangle and the box anomaly contribute to the total amplitude in a well defined way
\begin{equation}
A_{\gamma\to 3\pi}^{tot}=\frac{3}{2}A^{AVV}-\frac{1}{2}A^{VAAA}, 
\end{equation}   
where $A_{\gamma\to 3\pi}^{tot}$, $A^{AVV}$ and $A^{VAAA}$ are, respectively, the total $\gamma\pi\pi\pi$  amplitude, the point $\gamma\to\omega\to\pi\pi\pi$ amplitude and the amplitude for the $\gamma\to\omega\to\pi\rho\to\pi\pi\pi$ process. This result is consistent both with the chiral Ward identities and with the usual KSFR relation \cite{Kaw66,Riaz66}, which arises in the NJL model at $a=2$. Indeed, one can easily see from eq. (\ref{ff}) that, if one neglects the terms of order $1/a^2$ and higher in the box contribution and puts $c=0$ in the $\rho$-exchange term, the amplitude $A^{VAAA}$  has a factor $(1-3/a)=-1/2$, and the $A^{AVV}$ amplitude has a factor $(1-c/(2a)) 3/a=3/2$, as is required by the chiral Ward identities. On the other hand, if $c$ is chosen to cancel $\pi a_1$-mixing effects, these amplitudes contribute with a relative weight of $-7/64$ and $71/64$, correspondingly. 

Thus, the surface term $c$ cannot be used to resolve the $\pi a_1$-mixing puzzle. Moreover, its value is unambiguously fixed by the chiral Ward identities, which require that $c=0$. Exactly this pattern has been considered in \cite{Schechter84,Kaiser90,Wakamatsu89}, which reproduces well the phenomenological value of the width. That allows us to conclude, following \cite{Wakamatsu89}, that if the VMD is a valid theoretical hypothesis, the $\gamma\to\omega\to 3\pi$ amplitude contains the contributions due to $\pi a_1$-mixing and as a consequence the low energy theorem (\ref{leth}) is violated        
\begin{equation}
\label{g3pi}
A_{\gamma\to 3\pi}=-F^{3\pi} e_{\mu\nu\alpha\beta} \epsilon^\mu (q) p_0^\nu p_+^\alpha p_-^\beta ,
\end{equation}
\begin{equation}
\label{3pi}
F^{3\pi}=\frac{N_c e}{12\pi^2 f_\pi^3}\left(1+\frac{3}{2a^2}+\frac{1}{8a^3}\right)\neq \frac{N_c e}{12\pi^2 f_\pi^3}.
\end{equation}
The impossibility of ensuring the fulfilment of the low energy theorem without violating the so-called {\it complete} VMD was known earlier \cite{Fujiwara85,Bando88,Meissner88}. The peculiarity of the case under consideration is that the violation (\ref{3pi}) is associated with the presence of $\pi a_1$-transitions, i.e. occurs only when axial-vector mesons are present in the theory. In the following we will show that it is possible to combine the phenomenologically successful value $c=0$ with a full cancellation of $\pi a_1$-mixing effects within the NJL approach.

\section{The $\pi a_1$-mixing and $\gamma\to 3\pi$ amplitude}
To make further progress, let us recall that the $\pi a_1$ diagonalization is generally performed by a linearized transformation of the axial vector field. In the NJL model it has the following form 
\begin{equation}
\label{ngcr}
a_\mu \to a_\mu + \frac{\partial_\mu \pi}{ag_\rho f_\pi}, 
\end{equation}
where $\pi =\tau_i \pi^i$, $a_\mu =\tau_i a^i_\mu$ and $\tau_i$ are the $SU(2)$ Pauli matrices. It was this replacement that has been used in our calculations above. 

It is known however that the gauge noncovariant replacement (\ref{ngcr}) leads to the violation of gauge symmetry. The anomalous low energy amplitude describing the $a_1\to\gamma\pi^+\pi^-$ decay is not transverse \cite{Osipov18a,Osipov18b}. It has been also argued that gauge symmetry of the $a_1\to\gamma\pi^+\pi^-$ amplitude can be restored if one uses the covariant derivative $\mathcal D_\mu \pi$ instead of the gauge noncovariant one $\partial_\mu \pi$ 
\begin{equation}
\label{cov}
a_\mu \to a_\mu + \frac{\mathcal D_\mu \pi}{ag_\rho f_\pi}, \quad \mathcal D_\mu \pi =\partial_\mu \pi -ieA_\mu [Q,\pi ].
\end{equation}

\begin{figure}
\resizebox{0.20\textwidth}{!}{%
  \includegraphics{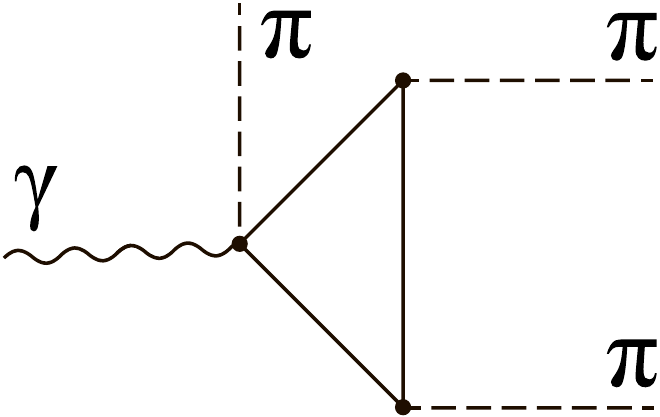}
}
\caption{The quark-loop graph contributing to the $\gamma\to 3\pi$ decay in the NJL model with the covariant $\pi a_1$ diagonalization (\ref{cov}). Both single pion lines are the result of $\pi a_1$-mixing. The graphs without $\pi a_1$-mixing on the single pion lines vanish.}
\label{fig3}      
\end{figure}

This modification of the theory does not affect any of the usual current-algebra theorems which involve amplitudes independent of $e_{\mu\nu\alpha\beta}$, whereas, it is important for processes with breaking of the intrinsic parity. Since we are dealing with exactly such a case here, let us apply this idea to the calculation of the $\gamma\to 3\pi$ amplitude. For that one should take into account an additional diagram which contributes to the $\gamma\to 3\pi$ amplitude (see Fig.~\ref{fig3}). The $\bar qq \gamma\pi$-vertex of this diagram stems from $\mathcal D_\mu \pi$ and induces a deviation of the theory from the complete VMD. The graph has three $\pi a_1$-transitions. One can easily check that similar diagrams with one and without $\pi a_1$-mixing effects on the separate pion lines do not contribute. 
 
The amplitude $A$, obtained by evaluation of the anomalous AAA triangle shown in Fig.~\ref{fig3}, is
\begin{eqnarray}
A&=&\frac{N_c e}{4a^3f_\pi^3}  
\left\{p_-^\sigma [J_{\mu\nu\sigma} (p_0,p_-)-J_{\mu\sigma\nu} (p_-,p_0)] \right. \nonumber \\
&& \!\!\!\!\!\! + \left. p_+^\sigma [J_{\mu\nu\sigma} (p_0,p_+)-J_{\mu\sigma\nu} (p_+,p_0)] \right\} \epsilon^\mu (q)p_0^\nu . 
\end{eqnarray}
The low energy expansion of the loop integral $J_{\mu\nu\sigma}$ starts from a linear term
\begin{equation}
\label{exp}
J_{\mu\nu\sigma} (p_0,p_-)=\frac{1}{24\pi^2} e_{\mu\nu\sigma\rho}\left(p_0-p_-  -3 \upsilon \right)^\rho +\ldots
\end{equation}
Owing to the shift ambiguity related to the formal linear divergence of this integral, the result depends on the undetermined 4-vector $\upsilon_\rho$, which survives in the final expression
\begin{equation}
\label{amp}
A=-\frac{N_c e}{4\pi^2 f_\pi^3} e_{\mu\nu\sigma\rho}\epsilon^\mu (q)p_0^\nu (p_++p_-)^\sigma \left(\frac{\upsilon^\rho}{4a^3}\right)
\end{equation}
Notice that this is the complete result for this triangle diagram. Terms of quartic and higher order in momenta are well defined (actually they vanish), while the cubic term has an undetermined, local contribution as given by eq. (\ref{amp}). 

The 4-vector $\upsilon_\rho$ can be represented as a linear combination of three independent momenta characterizing the process, $\upsilon_\mu= b_1 q_\mu + b_2 (p_+ -p_-)_\mu + b_3 (p_+ + p_-)_\mu$, where only the second term survives upon substituting this form into (\ref{amp}). Thus, the graph shown in Fig.~\ref{fig3} gives an additional contribution $\Delta F^{3\pi}$ to the form factor $F^{3\pi}$   
\begin{equation}   
\label{g3pinew} 
\Delta F^{3\pi}=\frac{N_c e}{12\pi^2 f_\pi^3} \left(\frac{-3b_2}{2a^3}\right),
\end{equation}
where $b_2$ is dimensionless and as yet undetermined. To fix it we use the low-energy theorem (\ref{leth}). By requiring that the unwanted terms in (\ref{3pi}) vanish we find that
\begin{equation}
\label{stcon}
b_2=a+\frac{1}{12}=1.92.
\end{equation}
Thus, the solution of the $\pi a_1$-mixing problem in the $\gamma\to 3\pi$ amplitude can be associated with the surface contribution of the anomalous non-VMD diagram shown in Fig.~\ref{fig3}.    

\section{Conclusions}
The QCD-inspired NJL model with vector and axial-vector mesons and electromagnetic interactions has been used to resolve a long standing puzzle -- the breaking of the low-energy theorem (\ref{leth}) due to $\pi a_1$-mixing effects. The proposed solution includes a new important step -- the covariant (with respect to the electromagnetic gauge transformations) $\pi a_1$ diagonalization (\ref{cov}). Since the gauge covariant derivative involves the electromagnetic field, a direct interaction of the photon with a pseudoscalar meson and a quark-antiquark pair appears. It gives rise to a new triangle graph which is finite but contains a superficial linear divergence. Due to the linear divergence, shifting the integration momentum in the closed quark loop changes the value of the integral, so that there is an essential ambiguity used to satisfy the low-energy theorem (\ref{leth}). This mechanism is beyond the VMD framework and deserves to be further investigated in the future. 

We conclude by remarking that the $\gamma\to 3\pi$ amplitude is already the second example (after the $a_1\to\gamma\pi^+\pi^-$ decay \cite{Osipov18a,Osipov18b}) where surface terms associated to an anomalous AAA contribution arising within a gauge covariant description of $\pi a_1$-mixing allow us to meet the important symmetry requirements. At the core of both problems has been the non gauge invariant VAAA box amplitude ($T^F_{\mu\nu\sigma\lambda}$ in notations of \cite{Zee72}). In the NJL model with spin-1 mesons this vertex appears also in the context of $\pi a_1$-mixing. For instance, in the description of the $a_1\to\gamma\pi^+\pi^-$ decay this graph arises in the amplitude with two $\pi a_1$ transitions, i.e. at the level of two derivatives, and leads to gauge-symmetry breaking. In the $\gamma\to 3\pi$ amplitude it is an integral element of two types of problematic Feynman diagrams - with two and three $\pi a_1$ transitions. Fortunately, the problem, as we just showed, is solved due to the contribution of the surface term of the diagram Fig.~\ref{fig3}. Incidentally, it was the inclusion of surface contributions in Ward identities that made it possible to obtain a reliable field-theoretical picture in relating $\gamma\to 3\pi$ to $\pi^0\to\gamma\gamma$ through Ward identities. The Ward identities involved there are in fact not anomalous. However, the surface term, which is normally dropped in standard applications of current-algebra, cannot be dropped in the correct Ward identities for $\gamma\to 3\pi$, where we must keep terms to third order in momentum. The reason for that is well explained in \cite{Zee72}.

\begin{acknowledgements}
The authors acknowledge financial support from FCT through the grant CERN/FIS-COM/0035/2019, and the networking support by the COST Action CA16201. A.~A. Osipov would like to thank M.~K. Volkov for his interest in the work and useful discussions.
\end{acknowledgements}


\end{document}